\documentclass[aps,pra,showpacs,twocolumn,amsmath,amssymb,groupedaddress]{revtex4-1}

\usepackage{graphicx}

\begin{document}

\title{Observation of Optical Solitons and Abnormal Modulation Instability in Liquid Crystals with Negative Dielectric Anisotropy}


\author{Jing Wang$^1$, Zhenlei Ma$^1$, Junzhu Chen$^1$, Jinlong Liu$^{1,2}$, Zhuo Wang$^1$, Yiheng Li$^1$, Qi Guo$^{1,4}$, Wei Hu$^{1,5}$,
 and Li Xuan$^3$}
\affiliation{
$^1$Guangdong Province Key Laboratory of Nanophotonic Functional Materials and Devices, South China Normal University, Guangzhou 510006, China}
\affiliation{$^2$College of Science, South China Agricultural University, Guangzhou, 510642, China}
\affiliation{$^3$State Key Laboratory of Applied Optics, Changchun Institute of Optics, Fine Mechanics and Physics, Chinese Academy of Sciences, Changchun 130033, China}
\affiliation{$^4$Corresponding author: guoq@scnu.edu.cn}
\affiliation{$^5$Corresponding author: huwei@scnu.edu.cn}


\begin{abstract}
We investigate theoretically and experimentally the optical beam propagation in the nematic liquid crystal with negative dielectric anisotropy,  which is aligned homeotropically in a $80\mu m$-thickness planar cell in the presence of an externally voltage. It is predicted that the nonlocal nonlinearity of liquid crystal undergo an oscillatory response function with a negative nonlinear refractive index coefficient. We found that the oscillatory nonlocal nonlinearity can support stable bright solitons, which are observed in experiment. We also found that abnormal modulation instability occurs with infinity gain coefficient at a fixed spatial frequency, which is no depend on the beam intensity. We observed the modulation instability in the liquid crystal at a very low intensity ($0.26W/cm^2$), and the maximum gain frequency  were found kept unchange when beam power changes over 2-3 orders of magnitude.
\end{abstract}

\date{\today}


\maketitle

\section{Introduction}
Nonlocal solitons, solitons in nonlinear media with a nonlocal response, have been the subject of intense theoretical and experimental studies in many physical systems\cite{Snyder1997,Krolikowski2004}, including photorefractive crystals \cite{Mitchell1998}, nematic liquid crystals \cite{Conti2003,Conti2004,Peccianti2004, Hu2006}, lead glasses \cite{Rotschild2005, Alfassi2007}, liquids \cite{Dreischuh2006, Conti2009}, liquid-filled photonic crystal fibers \cite{Rosberg2007,Rasmussen2009}, atomic vapors \cite{Skupin2007}, and Bose-Einstein condensates \cite{Pedri2005, Tikhonenkov2008}. There are various types of nonlocal solitons, such as vortex solitons \cite{Briedis2005, Kartashov2007}, multipole solitons \cite{Rotschild2006,Xu2005}, and soliton clusters \cite{Buccoliero2007,Deng2007,DDeng2007,Izdebskaya2013}. Nonlocal solitons have a number of interesting properties, including large phase shift \cite{Shou2011}, attraction between two out-of-phase solitons \cite{Hu2006,Dreischuh2006}, a self-induced fractional Fourier transform \cite{Lu2008}, and suppression of the collapse instability \cite{Bang2002}.

In the nonlocal response in nonlinear media,  the refractive index change at a given point is determined not only by the light intensity at that point but also by the light intensity near that point, which can be described by a nonlocal response function. There are several types of nonlocal response in real nonlinear media, such as the zeroth-order modified Bessel nonlocal response in nematic liquid crystals \cite{Hu2006}, the logarithmic nonlocal response in lead glass \cite{Shou2009}, and the exponential-decay type nonlocal response in aqueous solution of rhodamine B \cite{Ghofraniha2007} and diluted India ink \cite{gao2014}. All these response functions are integrable functions with non-negative value.

Nikolov {\it et al.} found that quadratic solitons are equivalent to nonlocal solitons  on the basis of an analogy between parametric interaction and diffusive nonlocality \cite{Nikolov2003}. The nonlocal analogy was later  used to successfully describe pulse compression \cite{Bache2007, Bache2008}, localized X waves \cite{Larsen2006}, and modulational instability \cite{Wyller2007} in $\chi^{(2)}$ materials. Nikolov {\it et al.} showed that the second harmonic (SH), like the nonlinear refractive index forming a waveguide, traps the fundamental wave (FW). They found two types of response functions, the exponential-decay type and the sine-oscillatory type. On the basis of  the sine-oscillatory type response function, one can understand why Buryak and Kivshar found numerically that quadratic solitons radiate linear waves \cite{Nikolov2003, Buryak1995}. In 2012, Esbensen {\it et al.} further investigated quadratic solitons with a sine-oscillatory type response function and found a family of analytical solutions under the strongly nonlocal approximation \cite{Esbensen2012}. However, these soliton solutions, whether numerical or analytical, were all unstable. We have shown that boundary conditions can stabilize this type of nonlocal soliton with sine-oscillatory response functions \cite{Wang2014}.  In this letter, we will demonstrate that nematic liquid crystal with negative dielectric anisotropy is new media which has oscillatory response functions.

Nematicons, spatial optical solitons in nematic liquid crystals (NLC), have been the subject of intense theoretical and experimental studies over the past two decades \cite{Peccianti2012}. The pioneering work on nematicons was reported in 1993 by Braun {\it et al.} \cite{Braun1993}. They investigated the strong self-focusing of a laser beam in NLC in various geometries, from which they recognized the importance of molecular reorientation and anchoring at the boundaries. Subsequently, in 1998, Warenghem {\it et al.} observed the beam self-trapping in capillaries filled with dye-doped NLC \cite{Warenghem1998}. In the same year, Karpierz {\it et al.} observed the same phenomenon in planar cells with homeotropically aligned NLC \cite{Karpierz1998}. They lowered the required power of observing nematicons to milliwatt levels. In 2000, Peccianti {\it et al.} reported on nematicon formation in planar cells containing a NLC aligned homogeneously in the presence of an externally applied voltage \cite{Peccianti2000}. They extended the propagation length of nematicons for millimeter levels and found the adequate model for describing nematicon propagation.

The investigation on nematicon was sparse until Conti {\it et al.} found that the NLC with a pretilt angle induced by an external
low-frequency electric field is a kind of strongly nonlocal nonlinear medium and nematicons are a kind of accessible solitons \cite{Conti2003, Conti2004, Snyder1997}. They derived a simplified model and linked nematicons with quadratic solitons. The basic properties on nematicon have been revealed gradually ever since. Among others, we have to mention the interactions between two nematicons \cite{Peccianti2002, Hu2006, Hu2008, Skuse2008}. Recently, Piccardi {\it et al.} reported the dark nematicon formation in planar cells filled with dye-doped NLC aligned homeotropically, which can provide an effective negative nonlinearity \cite{Piccardi2011}. The negative nonlinearity is realized through the guest-host interaction.

In this letter, we observed the nematicon formation in planar cells containing a NLC  with negative dielectric anisotropy and positive optical anisotropy homeotropically aligned in the presence of an externally applied voltage. Following the method in \cite{Conti2003}, we got a simplified model with a negative Kerr coefficient and an oscillatory periodic response function, which can support bright nematicons. We outlined the connection between the simplified model and the equations describing quadratic solitons \cite{Buryak1995, Nikolov2003, Esbensen2012, Wang2014}.

\section{physical model and soliton}
We performed a series of experiments to observe the nematicon formation in NLC with negative dielectric anisotropy. The experimental setup is illustrated in Fig. 1(a). A light beam from a Verdi laser was focused by a 10X microscope objective and launched into a 80-$\mu m$-thick NLC cell. The configuration of the cell was shown in Fig. 1(b) and the cell was filled with the homeotropically aligned KY19-008 NLC, whose $n_\parallel=1.726$, $n_\perp=1.496$, average elastic constant $K=1\times10^{-11}N$, optical anisotropy $\epsilon^{op}_a=0.74106$, and dielectric anisotropy $\epsilon^{rf}_a=-5.3$. Owing to the negative dielectric anisotropy, the NLC molecules will try to adjust in a low-frequency applied electrical field in such a manner that the molecule axes turn perpendicular to the direction of the electric field \cite{Schiekel1971}. A microscope and a CCD camera were used to collect the light scattered above the cell during propagation.
\begin{figure}[tp]
\centerline{\includegraphics[width=8.0cm]{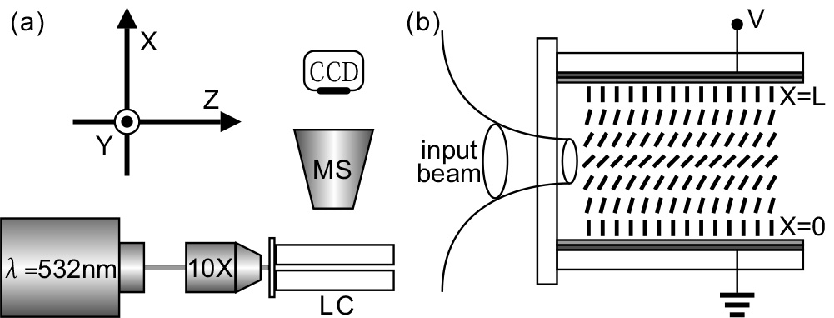}}
\caption{Sketch of the experimental setup (a) and homeotropically aligned nematic liquid crystal cell (b) for the observation of nematicons.}
\end{figure}

In the presence of an externally applied (low-frequency) electric field $E_{rf}$,  the evolution of the slowly varying envelope $A$ of a paraxial optical beam linearly polarized along $X$(an extraordinary light) and propagating along $Z$ can be described by the system,
\begin{equation}\label{eq2}
2ik\frac{\partial A}{\partial Z} + \nabla^2_{XY} A+k^2_0 \epsilon^{op}_a (\sin^2\theta-\sin^2\theta_0) A=0,
\end{equation}
\begin{equation}\label{eq3}
2K(\frac{\partial^2\theta}{\partial Z^2}+\nabla^2_{XY} \theta) +\epsilon_0(\epsilon^{rf}_a E^2_{rf}+\epsilon^{op}_a\frac{|A|^2}{2})\sin(2\theta) =0,
\end{equation}
where $\epsilon_0$ is the vacuum permittivity, $\theta$ is the tilt angle of the NLC molecules, $\theta_0$ is the nadir tilt in the absence of light, $k = k_0n_e(\theta_0)$ with $k_0$ the vacuum wavenumber and $n_e(\theta_0)=n_\perp n_\parallel/(n_\parallel^2cos^2\theta_0+n_\perp^2sin^2\theta_0)^{1/2}\approx(n_\perp^2+\epsilon^{op}_asin^2\theta_0)^{1/2}$ the refractive index of the extraordinary light at $\theta_0$, $\nabla^2_{XY}=\partial_X^2+\partial_Y^2$, $\epsilon^{rf}_a=\epsilon_\parallel-\epsilon_\perp(<0)$, $\epsilon^{op}_a=n_\parallel^2-n_\perp^2(>0)$. The term $\partial^2_Z\theta$ in Eq.(\ref{eq3}) was proven to be negligible compared to $\nabla^2_{XY} \theta$, therefore it can be removed. The homeotropical boundaries and anchoring at the interfaces define $\theta|_{X=0}=\theta|_{X=L}=\pi/2$, where $L$ is the cell thickness. In the absence of light, the pretilt angle $\hat{\theta}$ is symmetric along $X$ about $X = L/2$ (the cell center) and depends only on $X$:
\begin{equation}\label{eq4}
2K\frac{\partial^2\hat{\theta}}{\partial X^2} +\epsilon_0\epsilon^{rf}_a E^2_{rf}\sin(2\hat{\theta}) =0.
\end{equation}
As the external bias voltage increasing, the nadir tilt angle $\theta_0$ decreases when the voltage is above the Fr\'{e}edericks threshold.

Furthermore, analogous to the theory for nematicon in positive NLC\cite{Peccianti2004}, we can set $\theta=\hat{\theta}+(\hat{\theta}/\theta_0)\Phi$, with $\Phi$ being the optically induced perturbation. Noting that $\hat{\theta}\approx\theta_0$ and $\partial_X\hat{\theta}\approx0$ in the middle of the cell when the beam width $w_0$ is far smaller than the cell thickness, we can simplify Eq.(\ref{eq2}) and Eq.(\ref{eq3}) into the following system, which describes the coupling between $A$ and $\Phi$:
\begin{equation}\label{eq5}
2ik\frac{\partial A}{\partial Z} + \nabla^2_{XY} A+2n_0k^2_0 \Delta n A=0,
\end{equation}

\begin{equation}\label{eq6}
w^2_m\nabla^2_{XY}\Delta n +\Delta n =n_2|A|^2,
\end{equation}
where the nonlinear changes of diffraction index $\Delta n=\epsilon^{op}_a\sin(2\theta_0)\Phi/2n_0$, the parameter $w_m$ ($w_m>0$ for $|\theta_0| \leq \pi/2)$ is the characteristic length of the nonlinear response function, reads:
\begin{equation}\label{eq7}
w_m=\frac{1}{E_{rf}}\{\frac{2\theta_0K}{\epsilon_0|\epsilon^{rf}_a|\sin(2\theta_0)[1-2\theta_0\cot(2\theta_0)]}\}^{1/2}, \end{equation}
and \begin{equation}\label{eq8}
n_2=-\frac{(\epsilon^{op}_a)^2\theta_0\sin(2\theta_0)}{4n_0|\epsilon^{rf}_a|E_{rf}^2[1-2\theta_0\cot(2\theta_0)]}.
\end{equation}
The nonlinear refractive index coefficient $n_2$ is defined as suggested by Peccianti et al.\cite{Peccianti2004}.

The nonlocal nonlinearity in negative NLC described by Eq. (\ref{eq6}) is extraordinary than common nonlocal nonlinearity. First, the nonlinear refractive index coefficient  $n_2$ is negative for negative NLC,
while it is positive for  positive NLC. Secondly and more importantly, the response function
described by Eq. (\ref{eq6}) is oscillatory, given as (in 1+2D)
\begin{equation}\label{respfun2}
R(r)=Y_0(r/w_m)/(4w_m^2),
\end{equation}
or $R(x)=\sin(x/w_m)/(2w_m)$ (in 1+1D), where $r^2=x^2+y^2$, $Y_0(.)$ is zero-order Bessel function of the second kind. Then the nonlinear change of diffraction index $\Delta n$ is given as
\begin{equation}\label{dn}
\Delta n(r)=n_2\int_{-\infty}^\infty R(r-r^\prime)|A(r^\prime)|^2d^{D} r^\prime,
\end{equation}
where $\vec{r}=x\vec{i}+y\vec{j}$ for 1+2D(D=2) or $r=x$ for 1+1D(D=1). The response functions for negative NLC are not integrable, and it can not be normalized using  the condition  $\int_{-\infty}^\infty R(r)dr=1$. In local limit case (when $w_m$ approach zero), we have $\Delta n =n_2|A|^2$ from Eq.(\ref{eq6}) and nonlinearity is self-defocusing. But the response function  Eq.(\ref{respfun2}) can not approach a delta function mathematically.

It is very interesting to investigate asymptotic  behavior of these response functions for $r$ approach zero. From Eq.(\ref{respfun2}), we have $ R(r)\approx \ln(r/2w_m)/2\pi w_m^2$ for $r\rightarrow 0$, and it approach negative infinity. Considering that $n_2<0$ so the effective asymptotic function is  $ -\ln(r/2w_m)/2\pi w_m^2$, which is identical to the asymptotic function for positive NLC. For positive NLC, the response function is
$R(r)=K_0(r/w_m)/(2\pi w_m^2)$, where $K_0(.)$ is zero-order modified Bessel function of the second kind\cite{Peccianti2004,Hu2006}.  The same asymptotic functions for positive and negative NLC imply that under the strongly nonlocal approximation ($w_m\gg w_0$), the nonlinearity in negative NLC is self-focusing, although  the nonlinear refractive index coefficient  $n_2$ is negative. That means we can observe the bright soliton in negative NLC. Our numerical simulation and experimental observation confirmed our prediction as shown in Fig.2.

Numerical simulation of beam propagation were carried out based on Eq.(\ref{eq2}) and Eq.(\ref{eq3}) with Gaussian beam as an incident profile.  For low power input or without  the bias voltage, no nonlinear effect occurs and beams undergoe linear diffraction as shown in Fig.2(a). When increase beam power and bias voltage, we can found stable bright solitons as shown in Fig.2(b).

\begin{figure}[tp]
\centerline{\includegraphics[width=8.0cm]{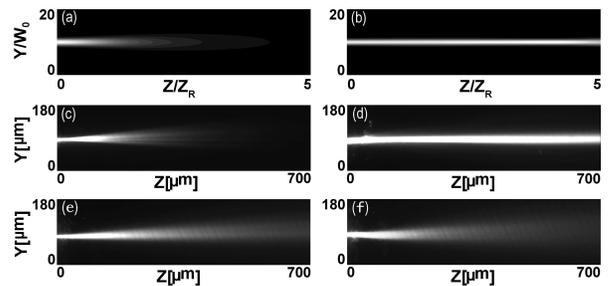}}
\caption{ (a)-(b) Numerical simulation of beam propagation based on Eq.(\ref{eq2}) and Eq.(\ref{eq3}) for  (a) linear diffraction and (b) soliton propagation. (c)-(d) $X$-polarized $e$-beam propagation  for (c) without voltage bias (linear diffraction) and (d) in the presence of 3.4 V at 1 kHz (soliton propagation). (e)-(f) $Y$-polarized $o$-beam propagation (linear diffraction) for (c) without voltage bias and (f) without voltage bias. Beam power $P=4.42mW$ and beam width $w=4\mu m$ in  (c)-(f).}
\end{figure}

In experiment, the launched power and beam width are fixed to $P_0=4.42 mW$ and $w_0=4\mu m$. For $x$-polarized beam ($e$-light), optically induced molecular reorientation occurs and contributes to the nonlinear index change. So soliton can be observed only for $x$-polarized beam.  Fig. 2(c) shows the linear diffraction in absence of the bias and Fig. 2(d) shows the nemation soliton formation at $V=3.4\text{V}$, which is  above  the Fr\'{e}edericks threshold. For  $y$-polarized beam ($o$-light), the nonlinearity does not occur and only diffraction beam can be observed, as shown in fig. 2(e) and 2(f).


\begin{figure}[tp]
\centerline{\includegraphics[width=8.0cm]{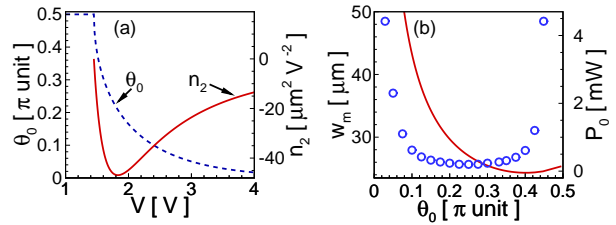}}
\caption{(Color online) (a) The pretilt angle $\theta_0$ and the Kerr coefficient $n_2$ of the NLC vs the bias voltage $V$. (b) The characteristic length $w_m$ (a solid curve) and the critical power of a single soliton (circles) vs the pretilt angle $\theta_0$. The parameters are for a $80-\mu m$-thick cell filled with the NLC (KY19-008) and the critical power is a numerical result.}
\end{figure}

The variations of characteristic length $w_m$ and  Kerr coefficient $n_2$ dependence on  pretilt angle $\theta_0$ and  the bias voltage $V$ are given in Fig.3.
A monotonous function of $\theta_0$ on $E_{rf}$ (or $V$ by $V=E_{rf}L$) is described by Eq. (\ref{eq4}). As shown in Fig. 3(a), $\theta_0$ decreases monotonously from $\pi/2$ to $0$ with increasing the bias above the Fr\'{e}edericks threshold. The Fr\'{e}edericks threshold
can be found \cite{Peccianti2004, Ruan2008}\begin{equation}\label{eq1}
V_{fr}=\pi(\frac{K}{\varepsilon_0|\varepsilon_a^{rf}|})^{1/2}.
\end{equation}
Introducing the value of $\epsilon^{rf}_a$ and $K$ for KY19-008 NLC, we can get $V_{fr}\approx1.45\text{V}$.
For $E_{rf}$ higher than the Fr\'{e}edericksz threshold, the approximation
\begin{equation}\label{eq9}
\theta_0\approx\frac{\pi}{2}(\frac{E_{fr}}{E_{rf}})^3
\end{equation}
is satisfactory, where $E_{fr}=V_{fr}/L$. Therefore, we can clearly see from Eqs.(\ref{eq7}) and (\ref{eq8}) that $w_m$ and $n_2$ are determined by $V$ or $\theta_0$ for a given NLC cell configuration. As shown in Fig. 3, $w_m$ and $n_2$ changes nonmonotonously with increasing $\theta_0$ and $V$, respectively. Both of them have a minimum value. There we also show the relation of the critical power $P_0$ on $\theta_0$ calculating numerically based on Eqs. (\ref{eq2})--(\ref{eq3}) with Gaussian beam as an incident profile.

\section{Modulation Instability}
The oscillatory-type response functions are singularity in the frequency spectrum \cite{Wang2014}. The response function
$R(r)$ in the Fourier domain $k_\perp$ is
\begin{equation}\label{eqfreq}
\tilde{R}(k_\perp)=1/(1-w_m^2k_\perp^2),
\end{equation}
where $k_\perp$ is transverse spatial frequency. It is singular when $k_\perp=1/w_m$. This singularity induce an abnormal MI in negative NLC.

Using the standard process, from Eq. \ref{eq5} and \ref{eq6} we find the MI occurs in negative NLC when
\begin{equation}
k_c^2<k_\perp^2<\frac{k_c^2}{2}+\frac{k_c^2}{2}\sqrt{\left(1+16n_0k_0^2|n_2|A_0^2\right)},
\end{equation}
where $k_c=1/w_m$ is the singular frequency, $A_0$ is the amplitude of background planar wave. The MI gain factor $g$ is given as
\begin{equation}\label{gain}
g(k_\perp)=\frac{|k_\perp|}{n_0k_0}\sqrt{\frac{(k_\perp^4-k_{c}^2k_\perp^2+4n_0k_0^2\gamma A_0^2k_{c}^2)}{(k_{c}^2-k_\perp^2)}}.
\end{equation}

\begin{figure}[tp]
\centerline{\includegraphics[width=8.0cm]{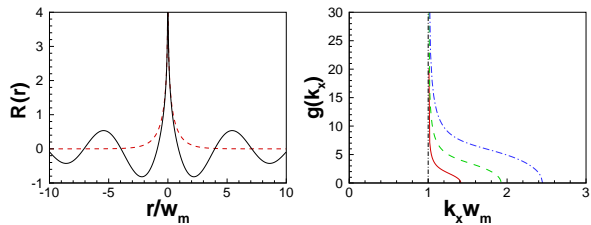}}
\caption{(Color online) (a) (1+2)D response function for negative (solid) and positive (dashed) NLC. (b) The modulation Instability gain for different beam power, $2n_0k_0^2n_2A_0^2=1, 5, 10$.}
\end{figure}

Figure 4(a) shows the profile of response functions for both negative and positive NLC. MI gain spectrum for different beam power was shown in Fig.4(b). We can see that at the singular frequency $k_c$, the MI gain factor is also singular and it value approach to infinity. When beam power increases, the MI range increase but the singular frequency keep unchange.  We can prediction that MI will occur at an extremely low power due to the infinity value of MI gain at the singular frequency $k_c$, which is independent on beam power.

We observed the abnormal MI in negative NLC using the same experimental configuration as Fig.1(a).  The $10X$-objective lens is replaced by a cylindrical lens with focal length $25mm$ to produce an ellipse beam with dimension of $10\mu m X 2.3mm$. The ellipse beam was launched in to the NLC cell and part of the beam was recorded by CCD, as shown in Fig.5. We have observed that MI occurs at low beam power of $0.06mW$, corresponding to beam intensity $0.26W/cm^2$. In our knowledge, it is the lowest intensity to induce MI in experiment.
\begin{figure}[tp]
\centerline{\includegraphics[width=8.0cm]{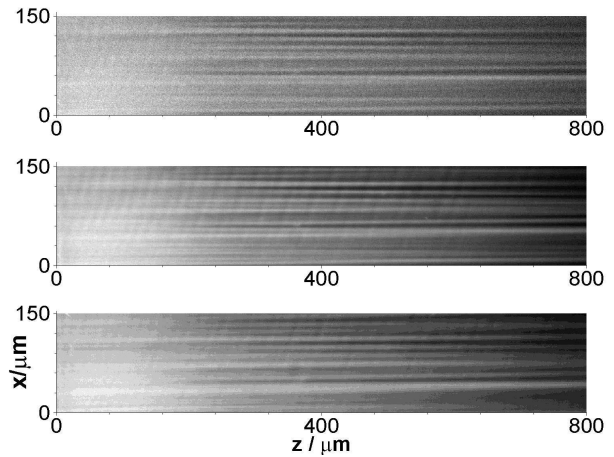}}
\caption{Experimental observation of MI for different power, $ P_0=0.06mW, 1.0mW, 100mW$, from top to bottom, respectively. The brightness of photo was increased for clear by software.}
\end{figure}

We read the intensity profiles from these photo and calculate the spatial spectrum intensities at different $z$-distance. By remove the initial spatial spectrum at $z=0\mu m$ we get the relative MI gain spectrum for several distance. The maximum MI gain frequencies $K_M$ are gotten and given in Fig.6(a).  One can see that the maximum MI gain frequencies fluctuate around $K_M\approx 0.05\mu m^{-1}$ when beam power changes from $0.06mw$ to $170mW$, over 3-order of magnitude. From our theory($k_c=1/w_m$), $K_M\approx 0.05\mu m^{-1}$ means the nonlinear characteristic length $w_m\approx 20\mu m$, which is near the value gotten from Eq. (\ref{eq7}). We also observed the maximum MI gain frequency for different bias voltage, as shown in Fig.6(a). At $z=400\mu m$, $K_M$ decreases as increasing of voltage. So we get the characteristic length $w_m$ decreases as increasing of $\theta_0$, same as shown in Fig.3(b). However, the value of $w_m$ gotten from  $K_M$ is always 3-4 times smaller than that in Fig.3(b). 

\begin{figure}[tp]
\centerline{\includegraphics[width=8.0cm]{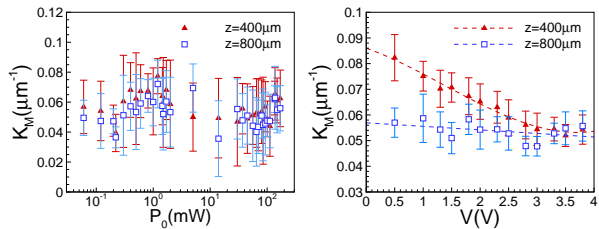}}
\caption{(a)Maximum MI gain frequency varies on the beam power, bias voltage $V=3.4V$. (b) Maximum MI gain frequency varies bias voltage.}
\end{figure}

\section{Conclusion}

In summary, we observed experimentally the nematicon formation in the planar cell containing the nematic liquid crystal with negative
dielectric anisotropy, aligned homeotropically in the presence of an externally applied voltage. We found that the nonlocal nonlinearity in this system undergo an oscillatory response function with a negative nonlinear refractive index coefficient.  The singularity of the oscillatory-type response functions induce an abnormal modulation instability, in which MI gain at singular frequency are infinity. In experiment,  we have observed that MI occurs at an extremely low beam intensity about $0.26W/cm^2$. So we demonstrate that negative NLC is new media with extraordinary oscillatory-type nonlocal nonlinearity.

The singularity of the oscillatory-type response function comes from the fact that the function extends to infinity in all space. If the oscillatory-type response function is limited within a finite space, the singularity will be suppressed \cite{Wang2014}. If we consider the response delay time, the oscillatory-type response function will be localized in space by a finite time. In this letter, the response delay is neglected, which is equivalence to that the oscillatory-type response function radiates for an infinity time. That is reason that MI gain approaches to infinity value.

This research was supported by the National Natural Science Foundation of China (Grant Nos. 11174090, 11174091, 11074080, and 11204299).

\end{document}